\documentclass[a4paper]{article}
\usepackage{multirow}
\usepackage{INTERSPEECH2022}

\usepackage[colorlinks,linkcolor=black,urlcolor=blue,citecolor=black]{hyperref}

\title{Speech Representation Disentanglement with Adversarial Mutual Information Learning for One-shot Voice Conversion
}

\name{Sicheng Yang$^{1,*}$, Methawee Tantrawenith$^{1,*}$, Haolin Zhuang$^{1,*}$\thanks{* Equal contribution}, Zhiyong Wu$^{1,3,\dagger}$\thanks{$\dagger$ Corresponding author}, Aolan Sun$^2$, \\
Jianzong Wang$^2$, Ning Cheng$^2$, 
Huaizhen Tang$^2$, Xintao Zhao$^1$, Jie Wang$^1$, Helen Meng$^{1,3}$}

\address{
  $^1$Shenzhen International Graduate School, Tsinghua University, Shenzhen, China\\
  $^2$Ping An Technology (Shenzhen) Co., Ltd., China\\
  $^3$The Chinese University of Hong Kong, Hong Kong SAR, China}
\email{\{yangsc21, tanmw20, zhuanghl21\}@mails.tsinghua.edu.cn,  zywu@sz.tsinghua.edu.cn
}

\begin{document}

\maketitle

\begin{abstract}
One-shot voice conversion (VC) with only a single target-speaker's speech for reference has become a hot research topic. 
Existing works generally disentangle timbre, while information about pitch, rhythm and content 
is still mixed together.
To perform one-shot VC effectively with further disentangling these speech components, we employ random resampling for pitch and content encoder and use 
the variational contrastive log-ratio upper bound of mutual information and gradient reversal layer based
adversarial mutual information learning to ensure the different parts of the latent space containing only the desired disentangled representation during training. 
Experiments on the VCTK dataset show the model achieves state-of-the-art performance for one-shot VC in terms of naturalness and intellgibility.
In addition, we can transfer 
characteristics
of one-shot VC on timbre, pitch and rhythm separately by speech representation disentanglement.
Our code, pre-trained models and demo are available at 
\href{https://im1eon.github.io/IS2022-SRDVC/}{https://im1eon.github.io/IS2022-SRDVC/}.
    
\end{abstract}

\section{Introduction}

Voice conversion (VC) is a speech task that studies how to convert one's voice to sound like that of another while preserving the linguistic content of the source speaker \cite{9262021, MOHAMMADI201765}.     
One-shot VC 
uses only one utterance from target speaker during the inference phase \cite{liu18d_interspeech, lu19_interspeech, Chou2019AdaIN, du2021improving}. Since one-shot VC requires small amount of data from the target speaker, it is more suitable with the needs of VC applications. Compare to traditional VC, to realize one-shot VC is more challenging. 


Many works for one-shot VC are based on speech representation disentanglement (SRD), which aim to separate speaker information from spoken content as much as possible.
The AutoVC \cite{qian2019autovc} comes up with the idea to combine the advantages of generative adversarial network (GAN) \cite{fang2018high} and conditional variational auto-encoder (CVAE) \cite{kameoka2018acvae} since GAN can obtain a good result while CVAE is easy to train. 
The IVC system and the SEVC system \cite{liu18d_interspeech}  represent the speaker
identity as i-vectors and speaker embedding to perform one-shot VC.
The unsupervised end-to-end automatic speech recognition and text-to-speech (ASR-TTS) autoencoder framework \cite{liu19c_interspeech} use multilabel-binary vectors to represent the content of speech to dientangle content from speaker.
An F0-conditioned voice conversion system \cite{9054734} disentangles prosodic features by tuning the information-constraining bottleneck of an autoencoder.
Due to the ability of SRD to disentangle latent space, there are many other methods \cite{Chou2019AdaIN, qian2019autovc, fang2018high, hsu2017voice, kaneko2020cyclegan, Wu2020VQVC} based on SRD for VC.

Speech information can be decomposed into four components: content, timbre, pitch and rhythm \cite{qian2020unsupervised}.
Rhythm characterizes the speed of the speaker uttering each syllable \cite{packman2000novel, gibbon2001measuring}.
Pitch is an important component of prosody \cite{qian2020unsupervised}.
Apparently, rhythm and pitch representations are related to content, which are important speech representations to improve the naturalness of converted speech \cite{helander2007novel}.
The above related works on VC only consider the decoupling of content and timbre representations without considering the rhythm and pitch representations related to the prosody of speech, which may lead to the leakage of information related to pitch and rhythm into the timbre
 \cite{9362110}. 

Recent studies begin to research pitch and rhythm in VC such as SpeechSplit \cite{qian2020unsupervised}, SpeechSplit2.0 \cite{9747763} and MAP Network \cite{wang2021adversarially}. They have provided a great start for the controllable synthesis of each representation.
Whereas, their performances can still be improved for one-shot VC.
A recent effort VQMIVC \cite{wang21n_interspeech} provides the pitch information to retain source intonation variations, resulting in high-quality one-shot VC.
However, it does not consider the decoupling of rhythm, 
resulting in that pitch and rhythm still entangles together.



In this paper, we propose adversarial mutual information learning based SRD for one-shot VC.
Specifically, we first decompose a speech into four factors: rhythm, pitch, content and timbre.
Then we propose a system consisting of four components:
(1) Random resampling \cite{8682589} operation along the time dimension is performed on pitch and content to remove rhythm information;
(2) A common classifier is used for timbre to extract features related to speaker identity, and a gradient reversal layer (GRL) based adversarial classifier is used for speaker-irrelevant information (content, pitch and rhythm) to eliminate speaker information in latent space;
(3) The variational contrastive log-ratio upper bound (vCLUB) of mutual information (MI) \cite{ChengHDLGC20} between different representations is used to separate speaker-irrelevant information from each other;
(4) A pitch decoder is used to reconstruct normalized pitch contour and a speech decoder is adopted to reconstruct mel-spectrogram.
Main contributions of our work are: 
(1) Implementing one-shot VC with decoupling content, timbre, rhythm and pitch from the speech,
which ensures pitch, rhythm and content more speaker-irrelevant and makes timbre more relevant to speaker.
(2) Applying SRD with vCLUB and GRL based adversarial mutual information learning without relying on text transcriptions.
Experiments show the information leakage issues can be effectively alleviated by applying with adversarial mutual information learning.


\section{Proposed Approach}

We aim to perform one-shot VC effectively 
by SRD.
In this section, we first describe our system architecture. Then we introduce the adversarial mutual information learning and show how one-shot VC on different speech representation is achieved.      

\subsection{Architecture of the proposed system}

\begin{figure}[t]
  \centering
  \includegraphics[width=0.86\linewidth]{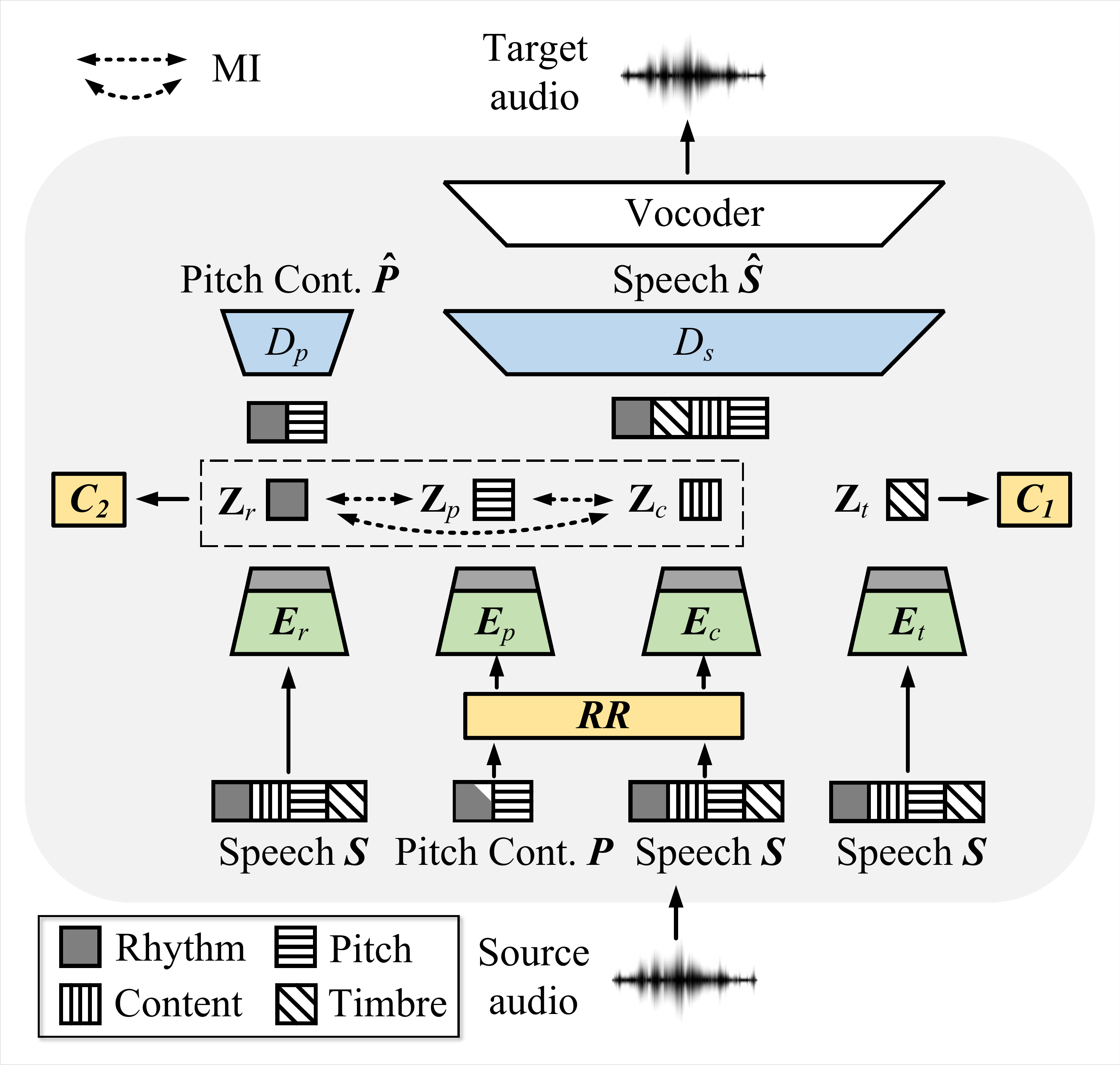}
  \caption{
  Framework of proposed model.
‘Pitch Cont.’ is short for the normalized pitch contour.  
Rhythm block with holes represents not all the rhythm information.
Signals are
represented as blocks to denote their information components.
}
  \label{fig:1}
\end{figure}

As shown in Fig.\ref{fig:1}, the architecture of the proposed model is composed of
six modules: 
rhythm encoder, pitch encoder, content encoder, timbre encoder, pitch decoder and speech decoder. 

\textbf{Speech representation encoders}: 
Inspired by SpeechSplit \cite{pmlr-v119-qian20a}, the rhythm encoder $E_r$, pitch encoder $E_p$, content encoder $E_c$ have the same structure, which
consists of a stack of 5 × 1 convolutional layers and a stack of bidirectional long short-term memory (BiLSTM) layers.
Besides mel-spectrogram, the pitch contour also carries information of rhythm \cite{qian2020unsupervised}.
Before the input mel-spectrogram $\boldsymbol{S}$ and normalized pitch contour $\boldsymbol{P}$ are fed to the content encoder and pitch encoder, random resampling \cite{8682589} operation along the time dimension is performed to remove rhythm information.
The model only relies on the rhythm encoder to recover the rhythm information.
The speaker encoder \cite{Chou2019AdaIN}  $E_t$ is used to provide global speech characteristics to control the speaker identity.
Denote the random resampling operation as $RR(\cdot)$, then we have:
\begin{equation}
\begin{split}
    \boldsymbol{Z}_{r}=E_{r}(\boldsymbol{S}), \quad \boldsymbol{Z}_{c}=E_{c}(RR(\boldsymbol{S})), \\
    \boldsymbol{Z}_{t}=E_{t}(\boldsymbol{S}), \quad
    \boldsymbol{Z}_{p}=E_{p}(RR(\boldsymbol{P})) \\
\end{split}
\end{equation}

\begin{figure}[t]
  \centering
  \includegraphics[width=\linewidth]{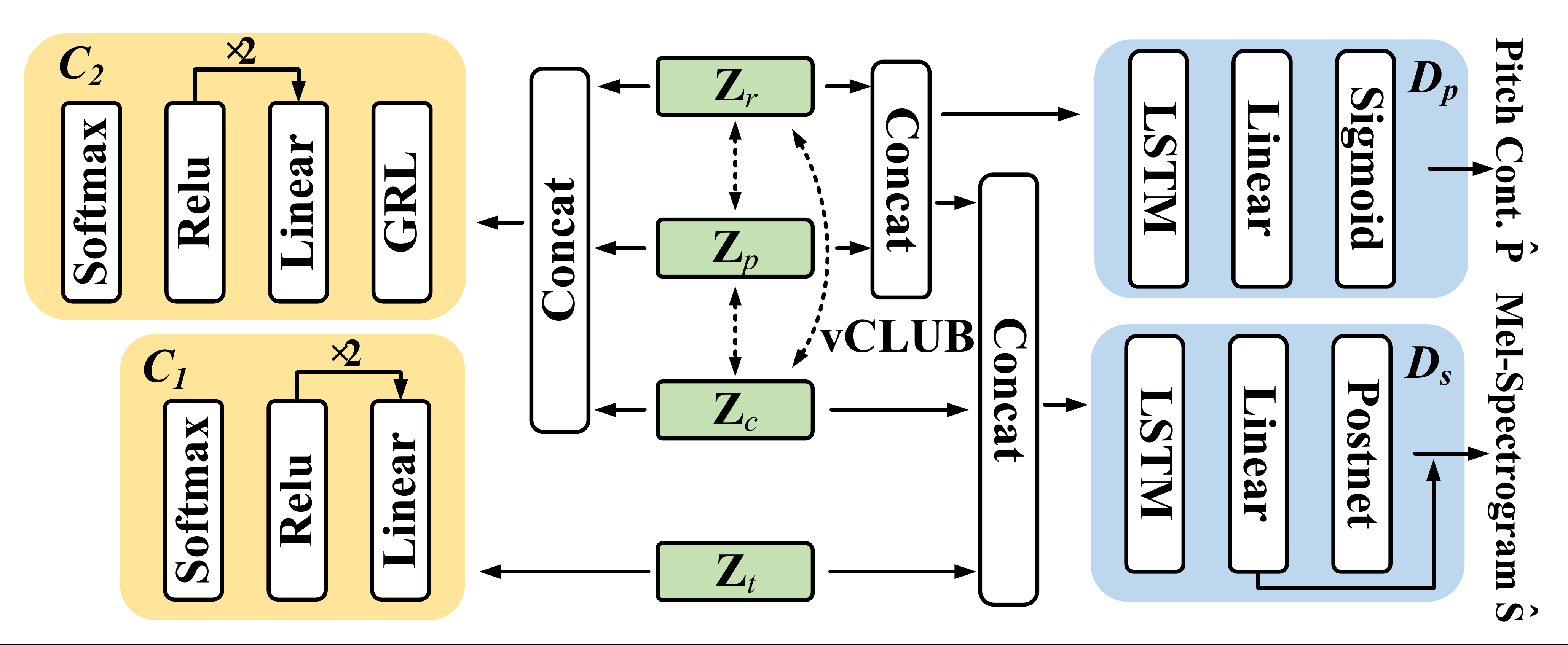}
  \caption{The architecture of proposed model. 
}
  \label{fig:2}
\end{figure}

\textbf{Decoders}: 
The speech decoder $D_s$ takes the output of all encoders as its input, and
outputs speech spectrogram $\hat{\boldsymbol{S}}$.
The input to the pitch decoder $D_p$ is $\boldsymbol{Z}_{p}$ and $\boldsymbol{Z}_{r}$. 
The output of $D_p$ is generated normalized pitch contour $\hat{\boldsymbol{P}}$.
As for implementation details, Fig.\ref{fig:2} shows the network
architecture used in our experiments.
\begin{equation}
    \hat{\boldsymbol{S}}=D_s\left(\boldsymbol{Z}_{r}, \boldsymbol{Z}_{p}, \boldsymbol{Z}_{c}, \boldsymbol{Z}_{t}\right), \quad \hat{\boldsymbol{P}}=D_p\left(\boldsymbol{Z}_{r}, \boldsymbol{Z}_{p}\right)
\end{equation}

During training, the output of $D_s$ tries to reconstruct
the input spectrogram $\boldsymbol{S}$, the output of $D_p$ tries to reconstruct
the input normalized pitch contour $\boldsymbol{P}$.
Decoders are jointly trained with encoders by minimizing the reconstruction losses:
\begin{equation}
    \mathcal{L}_{\text{s-recon}}=\mathbb{E}\left[\|\hat{\boldsymbol{S}}-\boldsymbol{S}\|_{1}^{2} + \|\hat{\boldsymbol{S}}-\boldsymbol{S}\|_{2}^{2}\right]
\end{equation}
\begin{equation}
 \mathcal{L}_{\text{p-recon}} = \mathbb{E}\left[\|\hat{\boldsymbol{P}}-\boldsymbol{P}\|_{2}^{2}\right]
\end{equation}




\subsection{Adversarial mutual information learning}

As shown in Fig.\ref{fig:1}, we use common classifier $C_1$ and adversarial speaker classifier $C_2$ to recognize the identity of speaker. 
The vCLUB \cite{ChengHDLGC20} is used to compute the upper bound of MI for irrelevant information of the speaker.
A measure of the dependence between two different variables can be formulated as:
\begin{equation}
\mathcal{I}(X, Y)=\int_{X} \int_{Y} P(X, Y) \log \frac{P(X, Y)}{P(X) P(Y)}
\end{equation}
where $P(X)$ and $P(Y)$ are the marginal distributions of $X$ and $Y$ respectively, and $P(X, Y)$ denotes the joint distribution of $X$ and $Y$. 

\textbf{Gradient Reverse}: 
Assumed the speaker information $u$ is 
related to timbre $\boldsymbol{Z}_{t}$ only, while independent of rhyme $\boldsymbol{Z}_{r}$, pitch $\boldsymbol{Z}_{p}$ and content $\boldsymbol{Z}_{c}$.
First, we aim to disentangle speaker-irrelevant information $\{\boldsymbol{Z}_{r}, \boldsymbol{Z}_{p}, \boldsymbol{Z}_{c}\}$ from speaker relevant timber $\boldsymbol{Z}_{t}$.
Instead of directly minimizing $\mathcal{I}(\{\boldsymbol{Z}_{r}, \boldsymbol{Z}_{p}, \boldsymbol{Z}_{c}\}, \boldsymbol{Z}_{t})$, we can maximize $\mathcal{I}(\boldsymbol{Z}_{t}, u)$
and minimize $\mathcal{I}(\{\boldsymbol{Z}_{r}, \boldsymbol{Z}_{p}, \boldsymbol{Z}_{c}\}, u)$ respectively,
which can be implemented as common classifier $C_1$ and adversarial speaker classifier $C_2$.
As shown in Fig.\ref{fig:2}, the input of $C_1$ is $\boldsymbol{Z}_{t}$ and the input of $C_2$ is $\{\boldsymbol{Z}_{r}, \boldsymbol{Z}_{p}, \boldsymbol{Z}_{c}\}$.
GRL \cite{2016Domain} is inserted in $C_2$ as the adversarial classifier. 
The two classifiers can be formulated as:
\begin{equation}
\begin{split}
    \mathcal{L}_{\text{com-cls}}\left(\boldsymbol{\theta}_{e_{c_{1}}}, \boldsymbol{\theta}_{c_{1}}\right)&=-\sum_{k=1}^{K} \mathbb{I}(u==k) \log p_{k}^{\prime} \\
\mathcal{L}_{\text {adv-cls }}\left(\boldsymbol{\theta}_{e_{c_{2}}}, \boldsymbol{\theta}_{c_{2}}\right)&=-\sum_{k=1}^{K} \mathbb{I}(u==k) \log p_{k}
\end{split}
\end{equation}
where $\mathbb{I}(\cdot)$ is the indicator function, $K$ is the number of
speakers 
and $u$ denotes speaker who produced speech $x$, $p_k$ is the probability of speaker $k$, $\theta_{e_{c_{1}}}$ denotes the parameters of $E_t$, and $\theta_{e_{c_{2}}}$ denotes the parameters of $\{E_r, E_p, E_c\}$.  
$\theta_{c_{1}}$ and $\theta_{c_{2}}$ denote the parameters of $C_1$ and $C_2$ respectively.

\textbf{Mutual information}: 
A vCLUB of mutual information is defined by:
\begin{equation}        
\begin{aligned}     
\mathcal{I}{(\boldsymbol{X} , \boldsymbol{Y})}=&  \mathbb{E}_{p(\boldsymbol{X}, \boldsymbol{Y})}\left[\log q_{\theta}(\boldsymbol{Y} \mid \boldsymbol{X})\right] \\
&-\mathbb{E}_{p(\boldsymbol{X})} \mathbb{E}_{p(\boldsymbol{Y})}\left[\log q_{\theta}(\boldsymbol{Y} \mid \boldsymbol{X})\right]
\end{aligned}
\end{equation}
where $\boldsymbol{X},\boldsymbol{Y} \in \{\boldsymbol{Z}_{r}, \boldsymbol{Z}_{p}, \boldsymbol{Z}_{c}\}$, $q_\theta(Y|X)$ is a variational distribution with parameter $\theta$ to approximate $p(Y|X)$. 
The unbiased estimator for vCLUB with samples $\left\{\boldsymbol{x}_{i}, \boldsymbol{y}_{i}\right\}$ is:
\begin{equation}        
\label{eq:1}
\hat{\mathcal{I}}(\boldsymbol{X} , \boldsymbol{Y})=\frac{1}{N^{2}} \sum_{i=1}^{N} \sum_{j=1}^{N}\left[\log q_{\theta}\left(\boldsymbol{y}_{i} \mid \boldsymbol{x}_{i}\right)-\log q_{\theta}\left(\boldsymbol{y}_{j} \mid \boldsymbol{x}_{i}\right)\right].
\end{equation}
where $\boldsymbol{x}_i,\boldsymbol{y}_i \in \{\boldsymbol{Z}_{r_{i}}, \boldsymbol{Z}_{p_{i}}, \boldsymbol{Z}_{c_{i}}\}$. 

By minimizing (\ref{eq:1}), we can decrease the correlation among different speaker-irrelevant speech representations. 
The MI loss is:
\begin{equation}
    \mathcal{L}_{MI} = \mathcal{\hat{I}}(\boldsymbol{Z}_{r}, \boldsymbol{Z}_{p}) + \mathcal{\hat{I}}(\boldsymbol{Z}_{r}, \boldsymbol{Z}_{c}) + \mathcal{\hat{I}}(\boldsymbol{Z}_{p}, \boldsymbol{Z}_{c})
\end{equation}

At each iteration of training, the variational approximation networks which are trained to maximize the log-likelihood $\log q_{\theta}(\boldsymbol{Y}|\boldsymbol{X})$ is first optimized, and then the VC network is optimized.
The loss of VC network can be computed as:
\begin{equation}
\label{loss}
    \mathcal{L}_{VC} = \mathcal{L}_{\text{s-recon}} + \mathcal{L}_{\text{p-recon}} + 
    \alpha\mathcal{L}_{\text{com-cls}} + 
    \beta\mathcal{L}_{\text {adv-cls}} + 
    \gamma\mathcal{L}_{MI}
\end{equation}

\subsection{One-shot VC on different speech representations}

Take general voice conversion (timbre conversion) as example. Rhythm $\boldsymbol{Z}_{r_\text{src}}$, pitch $\boldsymbol{Z}_{p_\text{src}}$ and content $\boldsymbol{Z}_{c_\text{src}}$ representations are
extracted from the source speaker’s speech $\boldsymbol{S}_\text{src}$. 
While timbre $\boldsymbol{Z}_{t_\text{tgt}}$ is extracted from the target speaker’s speech $\boldsymbol{S}_\text{tgt}$. 
The decoder then generates the converted
mel-spectrograms as $D_{s}(\boldsymbol{Z}_{r_\text{src}}, \boldsymbol{Z}_{p_\text{src}}, \boldsymbol{Z}_{c_\text{src}}, \boldsymbol{Z}_{t_\text{tgt}})$.
To convert rhythm, we feed
the target speech to the rhythm encoder $E_r$ to get the $\boldsymbol{Z}_{r_\text{tgt}}$. 
If we want to convert pitch, we feed
the normalized pitch contour of the target speaker to the pitch encoder $E_p$. 
Besides, three
double-aspect conversions (rhythm+pitch, rhythm+timbre,
and pitch+timbre) and all-aspect conversion are similar.

\section{Experiments}

\subsection{Experiment setup}
 
All experiments are conducted on the VCTK corpus \cite{VCTK}, which are randomly split into 100, 3 and 6
speakers as training, validation and testing sets respectively.
For acoustic features extraction, all audio recordings are downsampled to 16kHz, and
the mel-spectrograms are computed through a short-time
Fourier transform with Hann windowing, 
1024 for FFT size, 1024 for window size and 256 for hop
size. 
The STFT magnitude is transformed to mel scale
using 80 channel mel filter bank spanning 90 Hz to 7.6 kHz.
For pitch contour, z-normalization is performed for each utterance independently.

The proposed VC network is trained using the ADAM optimizer \cite{kingma2014adam} (learning rate is e-4, $\beta_1 = 0.9$, $\beta_2 = 0.98$) with a batch size of 16 for 800k steps. 
We set $\alpha=0.1, \beta=0.1, \gamma=0.01$ for equation (\ref{loss}) and use a pretrained
WaveNet \cite{van2016wavenet} vocoder to convert the output mel-spectrogram back to the waveform.
Due to the lack of explicit labels of the speech components, the degree of SRD are hard to evaluate \cite{pmlr-v119-qian20a}.
Therefore, we focus on the timbre conversion by comparing with other one-shot VC models (Please note that we are able to transfer four representations of the one-shot VC as the demo shows, not just the timbre of target speaker).
We compare our proposed method
 with AutoVC \cite{qian2019autovc}, ClsVC\cite{ClsVC}, AdaIN-VC \cite{Chou2019AdaIN}, VQVC+ \cite{Wu2020VQVC} and VQMIVC \cite{wang21n_interspeech}, which are among the
state-of-the-art one-shot VC methods.

\subsection{Experimental results and analysis}

\subsubsection{Conversion Visualization}

\begin{figure}[t]
  \centering
  \includegraphics[width=\linewidth]{ 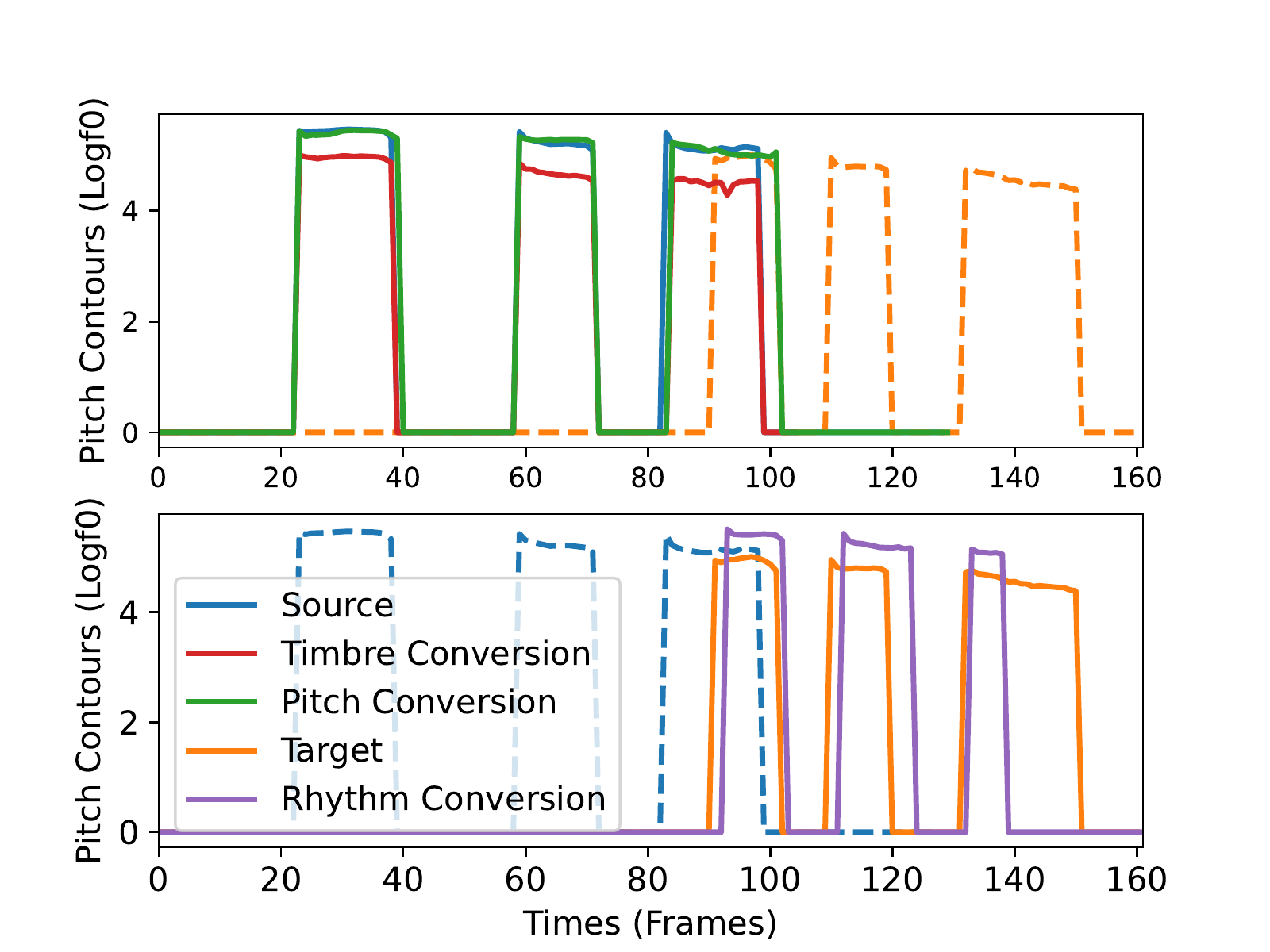}
  \caption{Pitch contours without normalization of single-aspect conversion results with the content ‘Please call Stella’.}
  \label{fig:4}
\end{figure}

%
Fig.\ref{fig:4} shows the pitch contours of the source (female speaker), target (male speaker) and converted speeches with the content
``Please call Stella''. 
Please note we use parallel speech data to visualize the results.
For timbre conversion, the pitch contour of the converted speech matches average pitch of the target speech but retains detailed characteristics of the source pitch contour.
For pitch conversion, on the other hand, the converted pitch contour tries to mimic detailed characteristics of the target speech with average pitch value close to the source one.
For rhythm conversion, the converted pitch contour is located at the target position as expected.
Furthermore, the timber conversion result in Fig.\ref{fig:4} also confirms to the common sense that the average pitch of the converted speech decreases from female to male.

\subsubsection{Subjective Evaluation}


Subjective tests are conducted by 32 listeners with good English proficiency to evaluate the speech naturalness and speaker similarity of converted speeches generated from different models.
The speech audios are presented to subjects in random order.
The subjects are asked to rate the speeches on a scale from 1 to 5 with 1 point interval.
The mean opinion scores (MOS) on naturalness and speaker similarity are reported in the first two rows 
in Table \ref{tab:3}.
The proposed model outperforms all the baseline models in terms of speech naturalness, and has a comparable performance with VQMIVC in terms of speaker similarity.

\begin{table*}[th]
\centering
\caption{Evaluation results of different models. 
Speech naturalness and speaker similarity are results of MOS with 95\% confidence intervals. `*' denotes the proposed model.
`w/o' is short for `without' in ablation study. `$\mathcal{L}_{\text{CLS}}$' denotes the $\mathcal{L}_{\text{com-cls}}$ and $\mathcal{L}_{\text{adv-cls}}$.}
\label{tab:3}
\resizebox{\textwidth}{!}{
    \begin{tabular}{c|cccccc|ccc}
    \toprule
    Methods & AutoVC & ClsVC & AdaIN-VC & VQVC+ & VQMIVC & Our Model* & w/o $\mathcal{L}_{\text{p-recon}}$* & w/o $\mathcal{L}_{\text{MI}}$* & w/o $\mathcal{L}_{\text{CLS}}$* \\
    \midrule
    Naturalness & 2.42$\pm$0.13 & 2.82$\pm$0.14 & 2.38$\pm$0.15 & 2.68$\pm$0.13 & 3.70$\pm$0.13 & \textbf{3.82$\pm$0.13} & 3.59$\pm$0.13 & 3.47$\pm$0.13 & 1.81$\pm$0.18 \\
    Similarity & 2.80$\pm$0.19 & 3.19$\pm$0.18 & 2.95$\pm$0.21 & 3.08$\pm$0.22 & \textbf{3.61$\pm$0.19} & 3.59$\pm$0.20 & 3.24$\pm$0.17 & 3.21$\pm$0.14 & 2.55$\pm$0.19 \\
    MCD/(dB) & 6.87 & 6.76 & 7.95 & 8.25 & 5.46 & \textbf{5.23} & 6.45 & 5.56 & 9.54 \\
    CER/(\%) & 32.69 & 42.90 & 29.09 & 38.17 & 10.08 & \textbf{9.60} & 19.63 & 6.91 & 65.00 \\
    WER/(\%) & 47.09 & 63.35 & 39.56 & 52.43 & 16.99 & \textbf{14.81} & 28.15 & 13.35 & 92.96 \\
    log$F_0$ PCC & 0.656 & 0.700 & 0.715 & 0.612 & \textbf{0.829} & 0.768 & 0.759 & 0.764 & 0.672 \\
    \bottomrule
    \end{tabular}
}
\end{table*}

\subsubsection{Objective Evaluation}

For objective evaluation, we use mel-cepstrum distortion (MCD) \cite{4317579}, character error rate (CER), word error rate (WER) and pearson correlation coefficient (PCC) \cite{Nahler2009Pearson} of log$F_0$.
MCD measures the spectral distance between two audio segments.
The lower the MCD is, the smaller the distortion, meaning that the two audio segments are more similar to each other.
To get the alignment between the prediction and the reference, we use dynamic time warping (DTW). 
CER and WER of the converted speech evaluate whether the converted voice maintains linguistic content of the source voice.
CER and WER are calculated by the transformer-based automatic speech recognition (ASR) model trained on the librispeech \cite{7178964}, which is provided by ESPnet2 \cite{li2020espnet}.
To evaluate intonation variations of the converted voice, PCC between F0 of source and converted voice is calculated.
To evaluate the proposed method objectively, 50 conversion pairs are randomly selected.
The results for different methods are shown in Table \ref{tab:3}.
It can be seen that our proposed model outperforms baseline systems on MCD. 
And our model achieves the lowest CER and WER among all methods, which
shows 
the proposed method has better intelligibility with preserve the source linguistic content.
In addition, 
our model achieves a higher log$F_0$ PCC, 
which shows the ability of our model in transforming and preserving the detailed intonation variations from source speech to the converted one.
Log$F_0$ PCC of VQMIVC is higher because the pitch information is directly fed to the decoder, without passing through the random resampling operation and pitch encoder.

\begin{figure}[t]
  \centering
  \includegraphics[width=0.85\linewidth]{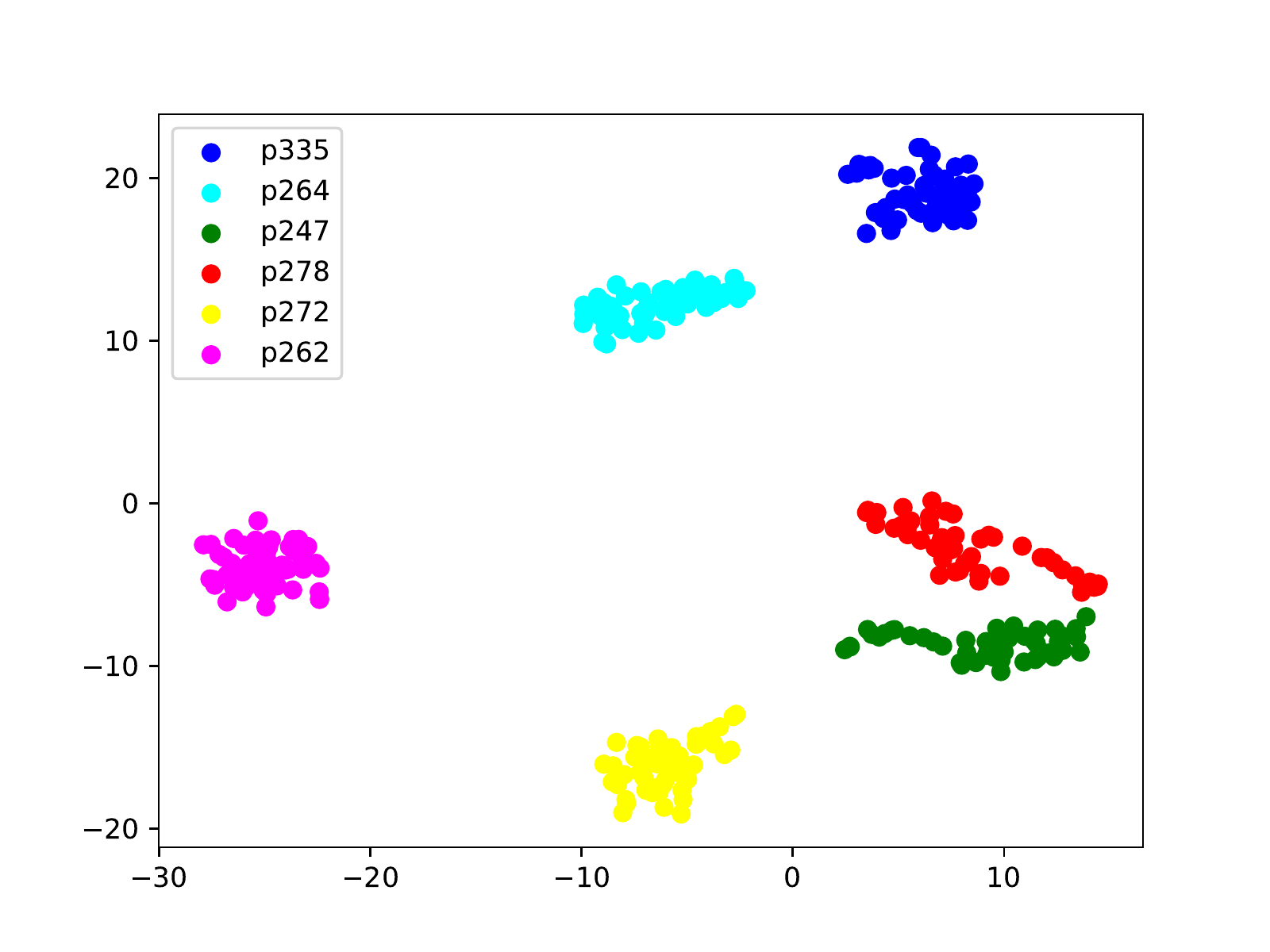}
  \caption{The visualization of speaker embedding. None of these speakers appeared in training.}
  \label{fig:3}
\end{figure}

Besides, Fig.\ref{fig:3} illustrates timber embedding $\boldsymbol{Z}_{t}$ of different speakers visualized by tSNE method.
There are 50 utterances sampled for every speaker to calculate the timbre representation. 
As can be seen, timber embeddings are separable for different speakers. 
In contrast, the timber embeddings of utterances of the same speaker are close to each other. 
The result indicates that our timber encoder $E_t$ is able to extract timbre $\boldsymbol{Z}_{t}$ as speaker information $u$.

\begin{figure}[t]
  \centering
  \includegraphics[width=0.85\linewidth]{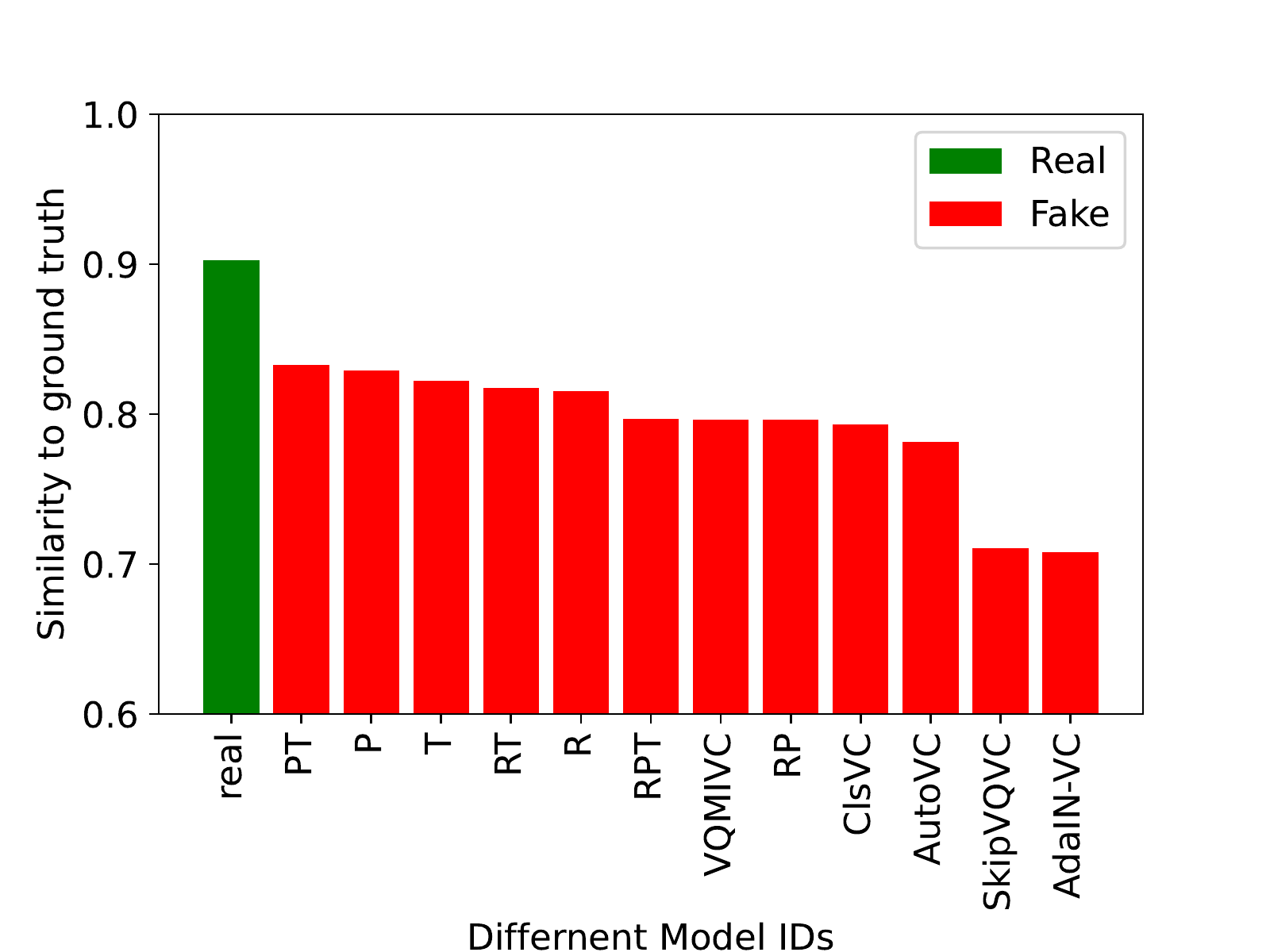}
  \caption{Results of fake speech detection. `P', `T' and `R' denotes pitch, timbre and rhythm conversion respectively.}
  \label{fig:5}
\end{figure}

To further evaluate the quality of one-shot VC on different speech representations, we use a  well-known open-source toolkit, Resemblyzer\footnote{\href{https://github.com/resemble-ai/Resemblyzer}{https://github.com/resemble-ai/Resemblyzer}},
to make the fake speech detection test.
On a scale of 0 to 1, the higher the score is, the more authentic the speech is.
50 utterances are used for each 
model.
The average scores are shown in Fig.\ref{fig:5}.
Our model performs well on one-shot VC on different speech representations.


\subsubsection{Ablation study}

Moreover, we conduct ablation study that addresses performance effects from different learning losses in equation (\ref{loss}), with results shown in the last three columns of Table \ref{tab:3}.
From the results, when the model is trained without the loss of pitch decoder $D_p$ ($\mathcal{L}_{\text{p-recon}}$) or vCLUB ($\mathcal{L}_{\text{MI}}$), the model is still able to perform one-shot VC and outperforms most of the baseline models, but the speech naturalness and speaker similarity both decrease.
When the losses of common classifier $C_1$ and adversarial speaker classifier $C_2$ ($\mathcal{L}_{\text{CLS}}$) are removed, the results 
are poor and no longer perform the VC task well.

\section{Conclusions}
In this paper, based on the disentanglement of different speech representations, we propose an approach using adversarial mutual information learning for one-shot VC. 
To make the timbre information as similar as possible to the speaker, we use a common classifier of the timbre.
Also, we use GRL to keep speaker-irrelevant information as separate from the speaker as possible.
Then pitch and content information can be removed from rhyme information by random resampling.
The pitch decoder ensures that the pitch encoder gets the correct pitch information.
The vCLUB makes speaker-irrelevant information as separate as possible.
We achieve proper disentanglement of rhythm, content, speaker and pitch representations and are able to transfer different representations style in one-shot VC separately.
The naturalness and intellgibility of one-shot VC is improved by speech representation disentanglement (SRD) and the performance and robustness of SRD is improved by adversarial mutual information learning.

\textbf{Acknowledgement}: 
This work is supported by National Natural Science Foundation of China (62076144), National Social Science Foundation of China (13\&ZD189) and Shenzhen Key Laboratory of next generation interactive media innovative technology (ZDSYS20210623092001004).


\clearpage

\bibliographystyle{IEEEtran}

\bibliography{main}

\end{document}